********************************************************************

\documentstyle[editedvolume]{crckapb} 

\input psfig.tex

\newcommand{\be}{\begin{equation}}
\newcommand{\ee}{\end{equation}}
\def\lsim{\lower.5ex\hbox{$\; \buildrel < \over \sim \;$}}
\def\gsim{\lower.5ex\hbox{$\; \buildrel > \over \sim \;$}}

\begin{opening}
\title{Dirac Equation in Kerr Geometry}

\author{Banibrata Mukhopadhyay}
\institute{Theoretical Astrophysics Group,\\
	S. N. Bose National Centre For Basic Sciences
	JD Block, Salt Lake, Sector-III, Calcutta-700091,
	India\\
	e-mail: bm@boson.bose.res.in}

\end{opening}

\runningtitle{Dirac Equation in Kerr Geometry }

\begin{document}

\begin{abstract}
{\small We are familiar with Dirac equation in flat space by which
we can investigate the behaviour of half-integral spin
particle. With the introduction of general
relativistic effects the form of the Dirac equation will
be modified. For the cases of different
background geometry like Kerr, Schwarzschild etc. the
corresponding form of the Dirac equation as well as
the solution will be different. In 1972, Teukolsky wrote the
Dirac equation in Kerr geometry. Chandrasekhar separated it
into radial and angular parts in 1976. Later Chakrabarti
solved the angular equation in 1984. In 1999 Mukhopadhyay
and Chakrabarti have solved the radial Dirac equation in
Kerr geometry in a spatially complete manner. In this
review we will discuss these developments systematically
and present some solutions.}
\end{abstract}
\smallskip
{\small
\noindent {\bf Keywords}~~: General relativity, spin-half particles, 
black holes, quantum aspects\\

\noindent {\bf PACS Nos.}~~:  04.20.-q, 04.70.-s, 04.70.Dy, 95.30.Sf}

\section{Introduction}

Behaviour of particles with half integral spin can be investigated through
the study of Dirac equation. Generally, we are familiar with the Dirac
equation and its solution when the space-time is flat. In the
curved space-time where the influence of the gravity is 
introduced, the corresponding equation
will be changed in form. Its solution will also be different. 
In 1972, Teukolsky [1] wrote the Dirac equation in curved space-time
particularly in Kerr geometry [2] using Newman-Penrose formalism [3]. 
Through this modified Dirac equation we can study the behaviour of spin half
particles around the spinning black holes. Due to presence of 
central black hole the space-time is influenced and behaviour of
the particle is changed with respect to that of flat space.
>From the same equation of Teukolsky, Dirac equation for
Schwarzschild metric [2] (Schwarzschild geometry), where the central
black hole is static can be studied just by putting the angular
momentum parameter $a$ of the black hole to zero. So one can
study how the behaviour of spin half particle in curved space
time is influenced by the angular momentum of black hole.
In 1976, Chandrasekhar [4] separated the Dirac equation in Kerr 
geometry into radial and angular
parts and solved the radial part of the equation asymptotically.
Chakrabarti in 1984 [5] solved the angular part
analytically. Here we shall introduce the spatially
complete analytical solution of radial Dirac equation [6-7]. So the
the complete solution of Dirac equation can be studied. Far
away from the black hole the the modified Dirac equation for curved
space-time (for Kerr and Schwarzschild geometry [2-3]) and its solution
reduce into that of the flat space.

In this review we will first indicate how Dirac equation in curved 
space-time can be written using Newman-Penrose formalism [3].
Newman-Penrose formalism is one of the tetrad formalism where
null basis are chosen instead of orthonormal basis. To fulfill
the understanding of Dirac equation in this formalism we also
need to know the `Spinor Analysis' [3]. In the next 
Section, we will briefly describe this in the context
of our present purpose. In \S 3 we will write the Dirac equation
in Newman-Penrose formalism for flat and curved space-time. 
For curved space we will separate the Dirac equation under
the background of Kerr geometry. In \S 4 and \S 5 we will briefly outline the 
angular and radial solution of Dirac equation respectively. In \S 6 we make concluding remarks.

\section{Spinor Analysis} 

In Minkowski space we consider a point $x^i$ ($i=0,1,2,3$)
on a null ray whose norm is defined as
$$
(x^0)^2-(x^1)^2-(x^2)^2-(x^3)^2=0.
\eqno{(1)}
$$
Now, we consider two complex numbers $\xi^0$ and $\xi^1$, and
their complex conjugate numbers ${\bar{\xi}}^{0'}$ and 
${\bar{\xi}}^{1'}$ in terms of which each point can be written as,
$$
x^0=\frac{1}{\sqrt{2}}(\xi^0{\bar{\xi}}^{0'}+\xi^1{\bar{\xi}}^{1'})
\eqno{(2a)}
$$
$$
x^1=\frac{1}{\sqrt{2}}(\xi^0{\bar{\xi}}^{1'}+\xi^1{\bar{\xi}}^{0'})
\eqno{(2b)}
$$
$$
x^2=\frac{1}{\sqrt{2}}(\xi^0{\bar{\xi}}^{1'}-\xi^1{\bar{\xi}}^{0'})
\eqno{(2c)}
$$
$$
x^3=-\frac{i}{\sqrt{2}}({\xi^0\bar{\xi}}^{0'}-\xi^1{\bar{\xi}}^{1'})
\eqno{(2d)}
$$
Conversely, we can write,
$$
\xi^0{\bar{\xi}}^{0'}=\frac{1}{\sqrt{2}}(x^0+x^3)
\eqno{(3a)}
$$
$$
\xi^0{\bar{\xi}}^{1'}=\frac{1}{\sqrt{2}}(x^1+ix^2)
\eqno{(3b)}
$$
$$
\xi^1{\bar{\xi}}^{0'}=\frac{1}{\sqrt{2}}(x^1-ix^2)
\eqno{(3c)}
$$
$$
\xi^1{\bar{\xi}}^{1'}=\frac{1}{\sqrt{2}}(x^0-x^3)
\eqno{(3d)}
$$
Let, 
$$
\xi_*^A={\alpha^A}_B{\xi^B},
\eqno{(4a)}
$$ 
$$
{\bar{\xi}}_*^{A'}={\bar{\alpha}}^{A'}_{B'}{\bar{\xi}}^{B'}
\eqno{(4b)}
$$
where, ($A,B,A',B'=0,1$), are the linear transformations in 
complex two-dimensional spaces. 
The transformation of $x^i$ is defined as,
$$
x^i_*=\beta^i_jx^j.
\eqno{(5)}
$$
Now, using equation (2) and (3) we can write,
$$
x^0_*=\frac{1}{\sqrt{2}}(\alpha^0_0\xi^0+\alpha^0_1\xi^1)
({\bar{\alpha}}^{0'}_{0'}{\bar{\xi}}^{0'}
+{\bar{\alpha}}^{0'}_{1'}{\bar{\xi}}^{1'})
+\frac{1}{\sqrt{2}}(\alpha^1_0\xi^0+\alpha^1_1\xi^1)
({\bar{\alpha}}^{1'}_{0'}{\bar{\xi}}^{0'}
+{\bar{\alpha}}^{1'}_{1'}{\bar{\xi}}^{1'})
$$
$$
=\frac{1}{2}(\alpha^0_0{\bar{\alpha}}^{0'}_{0'}+
\alpha^1_0{\bar{\alpha}}^{1'}_{0'})(x^0+x^3)
+\frac{1}{2}(\alpha^0_1{\bar{\alpha}}^{0'}_{1'}+
\alpha^1_1{\bar{\alpha}}^{1'}_{1'})(x^0-x^3)
$$
$$
+\frac{1}{2}(\alpha^0_0{\bar{\alpha}}^{0'}_{1'}+
\alpha^1_0{\bar{\alpha}}^{1'}_{1'})(x^1+ix^2)
+\frac{1}{2}(\alpha^0_1{\bar{\alpha}}^{0'}_{0'}+
\alpha^1_1{\bar{\alpha}}^{1'}_{0'})(x^1-ix^2) .
\eqno{(6)}
$$
Similarly, we can write down the relations between $x^1_*$, $x^2_*$
and $x^3_*$ with $\alpha$'s and $x$'s.
Therefore, keeping in mind (5) we can write,
$$
\beta^0_0+\beta^0_3=\alpha^0_0{\bar{\alpha}}^{0'}_{0'}+\alpha^1_0{\bar{\alpha}}^{1'}_{0'},
$$
$$
\beta^0_0-\beta^0_3=\alpha^0_1{\bar{\alpha}}^{0'}_{1'}+\alpha^1_1{\bar{\alpha}}^{1'}_{1'},
$$
$$
\beta^0_1-i\beta^0_2=\alpha^0_0{\bar{\alpha}}^{0'}_{1'}+\alpha^1_0{\bar{\alpha}}^{1'}_{1'},
$$
$$
\beta^0_1+i\beta^0_2=\alpha^0_1{\bar{\alpha}}^{0'}_{0'}+\alpha^1_1{\bar{\alpha}}^{1'}_{0'}.
$$ 
Now, imposing the condition that the transformation (5) is Lorentzian we can write,
$$
(\beta^0_0)^2-(\beta^0_1)^2-(\beta^0_2)^2-(\beta^0_3)^2=1
$$
So,
$$
\left|\left|\begin{array}{ccc}\alpha^0_0{\bar{\alpha}}^{0'}_{0'}
+\alpha^1_0{\bar{\alpha}}^{1'}_{0'}&{\alpha^0_0}{\bar{\alpha}}^{0'}_{1'}+
\alpha^1_0{\bar{\alpha}}^{1'}_{1'}\\{\alpha^0_1}{\bar{\alpha}}^{0'}_{0'}+
\alpha^1_1{\bar{\alpha}}^{1'}_{0'}&{\alpha^0_1}
{\bar{\alpha}}^{0'}_{1'}+\alpha^1_{1}{\bar{\alpha}}^{1'}_{1'}\end{array}
\right|\right| =1 .
\eqno{(7)}
$$
This gives,
$$
\Delta{\bar{\Delta}}=1
\eqno{(8)}
$$
Now we consider $\Delta = \bar{\Delta} =1$, so individually each transformation
of $\xi$ is Lorentzian. So we can conclude if transformation (5) is Lorentzian,
the necessary condition is transformation (4) is also Lorentzian.

Now we define spinors $\xi^A$, $\eta^{A'}$ of rank one as
$\xi_*^A=\alpha^A_B\xi^B$ and $\eta_*^{A'}={\bar{\alpha}}^{A'}_{B'}\eta^{B'}$,
($A, A', B, B' =0$), where $\left|\left|\alpha^A_B\right|\right|=
\left|\left|{\bar{\alpha}}^{A'}_{B'}\right|\right|=1$.
Since $\xi^A$ and $\eta^A$ are two spinors of same class,
$$
\left|\left|\begin{array}{ccc}\xi^0&\xi^1\\{\eta^0}&\eta^1\end{array}\right|
\right|=\xi^0\eta^1-\xi^1\eta^0 
\eqno{(9)}
$$
which is invariant under unimodular transformation, i.e.,
$$
\epsilon_{AB}\xi^A\eta^B \rightarrow {\rm invariant}
\eqno{(10)}
$$
where, $\epsilon_{AB}$ is Levi-Civita symbol. Here as in the case
of tensor analysis  $\epsilon_{AB}$ and  $\epsilon_{A'B'}$ are used
to lower the spinor indices as, $\xi_A=\xi^C\epsilon_{CA}$.

Now, using above information the representation of position vector
$x^i$ can be written as
$$
x^i \leftrightarrow \left|\begin{array}{ccc}\xi^0{\bar{\xi}}^{0'}&
\xi^0{\bar{\xi}}^{1'}\\{\xi^1}{\bar{\xi}}^{0'}&\xi^1{\bar{\xi}}^{1'}
\end{array} \right|=\frac{1}{\sqrt{2}}\left|\begin{array}{ccc}
x^0+x^3&x^1+ix^2\\x^1-ix^2&x^0-x^3\end{array}\right|
\eqno{(11)}
$$
Generally any vector $X^i$ can be written in terms of spinor
of rank two as,
$$X^i \leftrightarrow \left|\begin{array}{ccc}\xi^{00'}&\xi^{01'}\\
\xi^{10'}&\xi^{11'}\end{array}\right|=\frac{1}{\sqrt{2}}
\left|\begin{array}{ccc}X^0+X^3&X^1+iX^2\\X^1-iX^2&X^0-X^3\end{array}\right|
=X^{AB'}
\eqno{(12)}
$$
So a 4-vector is associated with a hermitian matrix such that,
$$
(X^0)^2-(X^1)^2-(X^2)^2-(X^3)^2=(X^0+X^3)(X^0-X^3)-(X^1+iX^2)(X^1-iX^2)
$$
$$
=2(\xi^{00'}\xi^{11'}-\xi^{01'}\xi_{10'})
=(\xi^{00'}\xi_{00'}+\xi_{11'}\xi^{11'}+\xi_{10'}\xi^{10'}+\xi^{01'}\xi_{01'})
=X_{AB'}X^{AB'}.
$$
>From the definition of norms, we can write it in two different representations:
$$
g_{ij}X^iX^j=\epsilon_{AC}\epsilon_{B'D'}X^{AB'}X^{CD'}
\eqno{(13)}
$$
Therefore, we can transform $X^i \leftrightarrow X^{AB'}$ using,
$$
X^i=\sigma^i_{AB'}X^{AB'}
\eqno{(14a)}
$$
$$
X^{AB'}=\sigma^{AB'}_{i}X^{i}
\eqno{(14b)}
$$
where, $\sigma^{AB'}_{i}$ and $\sigma^i_{AB'}$ are nothing but
Pauli matrices and their conjugate matrices with a factor $\frac{1}{\sqrt{2}}$.

A curved space-time is locally Minkowskian. 
At each point of space-time an orthonormal Dyad basis can
be set up as $\zeta_{(a)}^A$ and $\zeta_{({a'})}^{A'}$ ($a, {a'} = 0,1$
and $A, {A'} = 0,1$) for spinors.

We define, $\zeta_{(0)}^A=o^A$ and $\zeta_{(1)}^A=l^A$. The condition of
orthonormality is $\epsilon_{AB}o^Al^B=o^0l^1-o^1l^0=o_Bl^B=-o^Al_A=1$.

Also it is clear that, $\epsilon^{(a)(b)}\zeta_{(a)}^A\zeta_{(b)}^B
=\epsilon^{AB}$.

Then the null vectors are defined as

$l^i\leftrightarrow o^A{\bar{o}}^{B'}$, $m^i\leftrightarrow o^A{\bar{l}}^{B'}$,
${\bar{m}}^i\leftrightarrow l^A{\bar{o}}^{B'}$,
$n^i\leftrightarrow l^A{\bar{l}}^{B'}$.

Where, vectors obey relations of null tetrads such as,

$l^in_i=1, m^i{\bar{m}}_i=-1$ and all other products give zero.

In this way using dyad basis we can set up four null vectors which are
basis of Newman-Penrose formalism. Using (14) we can write the basis explicitly as
$$
l^i=\sigma^i_{AB'}\zeta_{(0)}^A{\bar{\zeta}}_{(0')}^{B'}=\sigma^i_{AB'}o^A{\bar{o}}^{B'},
\eqno{(15a)}
$$
$$
m^i=\sigma^i_{AB'}\zeta_{(0)}^A{\bar{\zeta}}_{(1')}^{B'}=\sigma^i_{AB'}o^A{\bar{l}}^{B'},
\eqno{(15b)}
$$
$$
{\bar{m}}^i=\sigma^i_{AB'}\zeta_{(1)}^A{\bar{\zeta}}_{(0')}^{B'}=
\sigma^i_{AB'}l^A{\bar{o}}^{B'},
\eqno{(15c)}
$$
$$
n^i=\sigma^i_{AB'}\zeta_{(1)}^A{\bar{\zeta}}_{(1')}^{B'}=\sigma^i_{AB'}l^A{\bar{l}}^{B'}.
\eqno{(15d)}
$$
Thus, in Newman-Penrose formalism the Pauli matrices change their forms as,
$$
\sigma^i_{AB'}=\frac{1}{\sqrt{2}}\left|\begin{array}{ccc}l^i&m^i\\{\bar{m}}^i&n^i
\end{array}\right|
\eqno{(16a)}
$$
$$
\sigma_i^{AB'}=\frac{1}{\sqrt{2}}\left|\begin{array}{ccc}n_i&-{\bar{m}}_i\\
-{\bar{m}}_i&l_i\end{array}\right|  .
\eqno{(16b)}
$$
Therefore in this basis, the directional derivatives can be written as, 

$D=l^i\partial_i$, ${\underline \Delta}=n^i\partial_i$, $\delta=m^i\partial_i$
and $\delta^*={\bar{m}}^i\partial_i$. 

Thus, the spinor equivalents of Newman-Penrose formalism are

$\partial_{00'}=D$, $\partial_{11'}={\underline \Delta}$, 
$\partial_{01'}=\delta$, $\partial_{10'}=\delta^*$.

Due to the reason, as explained earlier $\nabla_i\leftrightarrow \nabla_{AB'}$,
so we can write,
$$
\nabla_iX_j= X_{j;i}\leftrightarrow \nabla_{AB'}X_{CD'}=X_{CD';AB'},
$$
therefore,
$$
X_{CD';AB'}=\sigma^j_{CD'}\sigma^i_{AB'}X_{j;i}.
\eqno{(17)}
$$
For covariant derivatives spin coefficients $\Gamma$ are introduced. 
In the Newman-Penrose formalism these different
coefficients are assigned in terms of special symbols which 
are given in [3].

%

\section{Dirac Equation and its Separation}

Before going into discussion, we should mention about the unit of the
system. Here we have chosen throughout $h=c=G=1$, where $h=$ Plank
constant, $c=$ speed of light and $G=$ gravitational constant. It
is very clear that simultaneously all these quantities are chosen as
unity implying the corresponding system is dimensionless.

The Dirac equation in flat space using Newman-Penrose
formalism can be written as,
$$
\sigma^i_{AB'}\partial_iP^A+i\mu_*{\bar{Q}}_{B'}=0
\eqno{(18a)}
$$
$$
\sigma^i_{AB'}\partial_iQ^A+i\mu_*{\bar{P}}_{B'}=0 .
\eqno{(18b)}
$$
Here, $P^A$ and ${\bar{Q}}^{A'}$ are the pair of spinors.
$\mu_*/\sqrt{2}$ is the mass of the particles and $\sigma^i_{AB'}$
is nothing but Pauli matrix, because $1/\sqrt{2}$ factors
are canceled in the equation.

In curved space time Dirac equation reduces to 
$$
\sigma^i_{AB'}P^A_{;i}+i\mu_*{\bar{Q}}^{C'}\epsilon_{C'B'}=0,
\eqno{(19a)}
$$
$$
\sigma^i_{AB'}Q^A_{;i}+i\mu_*{\bar{P}}^{C'}\epsilon_{C'B'}=0 ,
\eqno{(19b)}
$$
where, $\sigma^i_{AB'}$ is same as defined in equation (16a).

Now, consider $B'=0$, then (19a) reduces to
$$
\sigma^i_{00'}P^0_{;i}+\sigma^i_{10'}P^1_{;i}-i\mu_*{\bar{Q}}^{1'}=0
$$
or,
$$
(\partial_{00'}P^0+\Gamma^0_{b00'}P^b)+(\partial_{10'}P^1+\Gamma^0_{b10'}P^b)
-i\mu_*{\bar{Q}}^{1'}=0,
$$
Therefore,
$$
(D+\Gamma_{1000'}-\Gamma_{0010'})P^0+(\delta^*+\Gamma_{1100'}
-\Gamma_{0110'})P^1-i\mu_*{\bar{Q}}^{1'}=0
\eqno{(20)}
$$
Similarly, choosing $B'=1$, we can get another similar type equation and
then we can get corresponding conjugate equation of both by interchanging
$P$ and $Q$. Now choosing,

$F_1=P^0$, $F_2=P^1$, $G_1={\bar{Q}}^{1'}$, $G_2=-{\bar{Q}}^{0'}$

and replacing various spin coefficients by their named symbols [3] we get the
the Dirac equation in Newman-Penrose formalism in its reduced form as,
$$
(D+\varepsilon-\rho)F_1+(\delta^*+\pi-\alpha)F_2=i\mu_*G_1,
\eqno{(21a)}
$$
$$
({\underline \Delta}+\mu-\gamma)F_2+(\delta+\beta-\tau)F_1=i\mu_*G_2,
\eqno{(21b)}
$$
$$
(D+\varepsilon^*-\rho^*)G_2-(\delta+\pi^*-\alpha^*)G_1=i\mu_*F_2,
\eqno{(21c)}
$$
$$
({\underline \Delta}+\mu^*-\gamma^*)G_1-(\delta^*+\beta^*-\tau^*)G_2=i\mu_*F_1.
\eqno{(21d)}
$$

\subsection{Basis Vectors of Newman-Penrose formalism in terms of Kerr Geometry}

The contravariant form of Kerr metric is given as [3],
$$
g^{ij}=\left(\begin{array}{cccc}\Sigma^2/\rho^2\Delta&0&0&2aMr/\rho^2\Delta\\
0&-\Delta/\rho^2&0&0\\0&0&-1/\rho^2&0\\2aMr/\rho^2\Delta&0&0&
-(\Delta-a^2sin^2\theta)/\rho^2{\Delta}sin^2\theta\end{array}\right)
\eqno{(22)}
$$
\noindent where, $E$ is the energy, $a$ is specific angular momentum of the black hole,
$M=$ mass of the black hole, $\rho^2=r^2+a^2cos^2\theta$ (should not
confuse with the spin coefficient $\Gamma_{(0)(0)(1)(0')}=\rho$),
$\Sigma^2=(r^2+a^2)^2-a^2\Delta sin^2\theta$, $\Delta=r^2+a^2-2Mr$.

In Kerr geometry, the tangent vectors of null geodesics are:
$\frac{dt}{d\tau}=\frac{(r^2+a^2)}{\Delta}E$, $\frac{dr}{d\tau}=\pm E$,
$\frac{d\theta}{d\tau}=0$, $\frac{d\phi}{d\tau}=\frac{a}{\Delta}E$, 
\noindent where $\tau$ is the proper time (not to be confused with spin coefficient
$\Gamma_{(0)(0)(1)(1')}=\tau$).

Now, the basis of Newman-Penrose formalism can be defined in Kerr
geometry as (in tetrad form),
$$
l_i=\frac{1}{\Delta}(\Delta, -\rho^2, 0, -a\Delta {\rm sin}^2\theta) ,
\eqno{(23a)}
$$
$$
n_i=\frac{1}{2\rho^2}(\Delta, \rho^2, 0, -a\Delta {\rm sin}^2\theta) ,
\eqno{(23b)}
$$
$$
m_i=\frac{1}{{\bar{\rho}}\sqrt{2}}(i {\rm asin}\theta, 0, -\rho^2, -i(r^2+a^2){\rm sin}\theta),
\eqno{(23c)}
$$
$$
l^i=\frac{1}{\Delta}(r^2+a^2, \Delta, 0, a),
\eqno{(23d)}
$$
$$
n^i=\frac{1}{{\bar{\rho}}\sqrt{2}}(r^2+a^2, -\Delta, 0, a),
\eqno{(23e)}
$$
$$
m_i=\frac{1}{{\bar{\rho}}\sqrt{2}}(i {\rm asin}\theta, 0, 1, i {\rm cosec}\theta),
\eqno{(23f)}
$$
${\bar{m}}_i$ and ${\bar{m}}^i$ are nothing but complex conjugates of
$m_i$ and $m^i$ respectively.

\subsection{Separation of Dirac Equation into Radial and Angular parts}

It is clear that the basis vectors basically become derivative operators
when these are applied as tangent vectors to the function $e^{i(\sigma t+m\phi)}$.
Here, $\sigma$ is the frequency of the particle (not to be confused with spin
coefficient $\Gamma_{(0)(0)(0)(1')}=\sigma$) and $m$ is the azimuthal
quantum number [3].

Therefore, we can write,

$\vec{l}=D={\cal D}_0$, $\vec{n}= {\underline \Delta}=
-\frac{\Delta}{2\rho^2}{\cal D}_0^{\dag}$, $\vec{m}=\delta=\frac{1}
{{\bar{\rho}}\sqrt{2}}{\cal L}_0^{\dag}$, $\vec{\bar m}=\delta^*=\frac{1}
{{\bar{\rho}}^*\sqrt{2}}{\cal L}_0$.

where,
$$
{\cal D}_n=\partial_r+\frac{iK}{\Delta}+2n\frac{r-M}{\Delta},
\eqno{(24a)}
$$
$$
{\cal D}_n^{\dag}=\partial_r-\frac{iK}{\Delta}+2n\frac{r-M}{\Delta},
\eqno{(24b)}
$$
$$
{\cal L}_n=\partial_\theta+Q+n {\rm cot}\theta
\eqno{(25a)}
$$
$$
{\cal L}_n^{\dag}=\partial_\theta-Q+n{\rm cot}\theta.
\eqno{(25b)}
$$
$K=(r^2+a^2)\sigma+am$, $Q=a\sigma {\rm sin}\theta+m {\rm cosec}\theta$.

The spin coefficients can be written as combination of basis vectors in 
Newman-Penrose formalism which are now expressed in terms of elements of
different components of Kerr metric. So we are combining 
those different components
of basis vectors in a suitable manner and get the spin coefficients as,
$$
\kappa=\sigma=\lambda=\nu=\varepsilon=0 .
\eqno{(26a)}
$$

${\tilde \rho}=-\frac{1}{{\bar{\rho}}^*}$, $\beta=\frac{cot\theta}{{\bar{\rho}}^*
2\sqrt{2}}$, $\pi=\frac{iasin\theta}{({\bar{\rho}}^*)^2\sqrt{2}}$,

$\tau=-\frac{iasin\theta}{\rho^2\sqrt{2}}$, $\mu=-\frac{\Delta}{2\rho^2
{\bar{\rho}}^*}$, $\gamma=\mu+\frac{r-M}{2\rho^2}$, $\alpha=\pi-\beta^*$.\hskip2.5cm
(26b)

Using the above definitions and results and choosing $f_1={\bar{\rho}}^*F_1$,
$g_2={\bar{\rho}}G_2$, $f_2=F_2$, $g_1=G_1$ the Dirac equation is reduced to
$$
{\cal D}_0f_1+2^{-1/2}{\cal L}_{1/2}f_2=(i\mu_*r+a\mu_*cos\theta)g_1,
\eqno{(27a)}
$$
$$
\Delta{\cal D}_{1/2}^{\dag}f_2-2^{1/2}{\cal L}_{1/2}^{\dag}f_1=
-2(i\mu_*r+a\mu_*cos\theta)g_2,
\eqno{(27b)}
$$
$$
{\cal D}_0g_2-2^{-1/2}{\cal L}_{1/2}^{\dag}g_1=(i\mu_*r-a\mu_*cos\theta)f_2,
\eqno{(27c)}
$$
$$
\Delta{\cal D}_{1/2}^{\dag}g_1+2^{1/2}{\cal L}_{1/2}g_2=-2(i\mu_*r-a\mu_*cos\theta)f_1,
\eqno{(27d)}
$$
Now we will separate the Dirac equation into radial and angular parts by choosing,

$f_1(r, \theta)=R_{-1/2}(r)S_{-1/2}(\theta)$, $f_2(r, \theta)=R_{1/2}(r)S_{1/2}(\theta)$,

$g_1(r, \theta)=R_{1/2}(r)S_{-1/2}(\theta)$, $g_2(r, \theta)=R_{-1/2}(r)S_{1/2}(\theta)$.

Replacing these $f_i$ and $g_i$ ($i=1,2$) into (27) and using separation constant 
$\lambda$ we get,
$$
{\cal L}_{1 \over 2} S_{+{1 \over 2}} = - (\lambda -a m_p \cos \theta) S_{- {1 \over 2}}
\eqno{(28a)}
$$
$$
{\cal L}_{1 \over 2}^{\dag} S_{-{1 \over 2}} = + (\lambda+a m_p \cos \theta) S_{+{1\over2}}
\eqno{(28b)}
$$
$$
\Delta^{\frac{1}{2}}{\cal D}_{0} R_{- \frac{1}{2}}
= ( \lambda + i m_p r) \Delta^{\frac{1}{2}} {R}_{+ \frac{1}{2}} ,
\eqno{(29a)}
$$
$$
\Delta^{\frac{1}{2}} {\cal D}_{0}^{\dag} \Delta^{1 \over 2}
{R}_{+{\frac{1}{2}}} = ( \lambda - i m_p  r)  {R}_{-{1 \over 2}} ,
\eqno{(29b)}
$$
where, $m_p$ is the mass of the particle which is nothing but $2^{1/2}\mu_*$. Also,
$2^{1/2}R_{-1/2}$ is redefined as $R_{-1/2}$.

Equations (28) and (29) are the angular and radial Dirac equation respectively 
in coupled form with the separation constant $\lambda$ [3].

\section{Solution of Angular Dirac Equation}

Decoupling equation (28) we obtain the eigenvalue equation for spin-$\frac{1}{2}$ 
particles as 
$$
\left [{\cal L}_{1 \over 2} {\cal L}_{1 \over 2}^{\dag} +
\frac{a m_p \sin \theta}{\lambda + a m_p \cos \theta} {\cal L}_{1\over 2}^{\dag} +
(\lambda^2 - a^2 m_p^2 \cos^2 \theta) \right ] S_{-{1 \over 2}} = 0 .
\eqno{(30)}
$$
Similarly, one can obtain decoupled equation for spin+$\frac{1}{2}$ particles.
Here, the separation constant $\lambda$ is considered to be the eigenvalue of the
equation. The exact solutions of this equation for $\lambda$ and $S_{-{1\over 2}}$
is possible in terms of orbital
angular momentum quantum number $l$ and the spin of the particle $s$
when the parameter $\rho_1=\frac{m_p}{\sigma}=1$. When the
angular momentum of the black hole is zero i.e., Schwarzschild case, 
the equation is reduced in such a form that whose solution is nothing but
standard spherical harmonics such as [8-9],
$$
S_{-1/2}(\theta)e^{im\phi}=_{-{1 \over 2}}\!Y_{lm}(\theta, \phi),
\eqno{(31)}
$$
the eigenvalue i.e., the separation constant can be solved as,
$$
\lambda^2=(l+1/2)^2 .
\eqno{(32)}
$$
Similarly, for spin+$\frac{1}{2}$ particle one can solve $S_{+1/2}$ as
$$
S_{+1/2}(\theta)e^{im\phi}=_{+{1 \over 2}}\!Y_{lm}(\theta, \phi),
\eqno{(33)}
$$
with same eigenvalue $\lambda$.

For any non-integral, massless, spin particle the solutions are [8-9]
$$
S_{\pm s}(\theta)e^{im\phi}=_{\pm s}\!Y_{lm}(\theta, \phi),
\eqno{(34)}
$$
$$
\lambda^2=(l+|s|)(l-|s|+1).
\eqno{(35)}
$$
In the case of Kerr geometry, when $a\neq 0$ the equation can be solved 
by perturbative procedure [5] with perturbative parameter
$a\sigma$. The solution for $\rho_1=m_p/\sigma=1$ and $s=\pm {1 \over 2}$ is [5]
$$ 
\lambda^2 = (l+\frac{1}{2})^2 + a \sigma ( p+ 2m) + a^2 \sigma^2 
\left [1-\frac{y^2}{2(l+1)+a\sigma x} \right ] ,
\eqno{(36)}
$$
$$
{}_{1\over 2}S_{lm} =
{}_{1\over 2}Y_{lm} + \frac{a\sigma y}{2(l+1)+a\sigma x} {}_{1\over 2}Y_{l+1 m}
\eqno{(37a)}
$$
$$
{}_{- {1\over 2}}S_{lm} =
{}_{-{1\over 2}}Y_{lm} - \frac{a\sigma y}{2(l+1)+a\sigma x} {}_{-{1\over 2}}Y_{l+1 m}
\eqno{(37b)}
$$
where,
$$
p=F(l,l); \ \ \ x=F(l+1,l+1); \ \ \ y=F(l,l+1)
\eqno{(38)}
$$
and
$$
F(l_1,l_2)=[(2l_2+1)(2l_1+1)]^{\frac{1}{2}} <l_2 1 m 0|l_1 m>
[<l_2 1 \frac{1}{2} 0|l_1 \frac{1}{2}> 
$$
$$
+(-1)^{l_2-l}<l_2 1 m 0|l_1 m>[<l_2 1 \frac{1}{2} 0|l_1 \frac{1}{2}> +(-1)^{l_2-l}
\rho_1 \sqrt{2} <l_2 1 -\frac{1}{2} 1|l_1 \frac{1}{2}>]] .
\eqno{(39)}
$$
with $<....|..>$ are the usual Clebsh-Gordon coefficients.

If $\rho_1 \neq 1$ then exact solution is not possible. In those cases the analytic
expression of eigenvalue and angular wave-function are found as infinite series
not in a compact form as the case $\rho_1=1$ . 

>From the general convergence of series expansions one can truncate the infinite series
upto certain order for particular values of $l$, $s$ and $m$. For $l=\frac{1}{2}$,
$s=-\frac{1}{2}$ and $m=-\frac{1}{2}$, up to third order in $a\sigma$, one obtains [5], 
$$
\lambda^2=(l+\frac{1}{2})^2+a\sigma f_1(l,m)+(a\sigma)^2 f_2(l,m)+(a\sigma)^3 f_3(l,m),
\eqno{(40)}
$$
$$
{}_{-\frac{1}{2}}S_{\frac{1}{2} -\frac{1}{2}}=-sin\theta-\left(sin^3\frac{\theta}{2}
-sin\theta cos\frac{\theta}{2}\right)\left[\frac{2}{3}a\sigma(1+\rho_1)+\frac{4}{15}
(a\sigma)^2(1-\rho_1^2)\right]
$$
$$
+\frac{2}{5}(a\sigma)^2(1-\rho_1^2)\left[sin^5\frac{\theta}{2}
-6sin^2\frac{\theta}{2}cos^3\frac{\theta}{2}+3sin\frac{\theta}{2}cos^4\frac{\theta}{2}\right] .
\eqno{(41)}
$$
The accuracy of eigenvalues and eigenfunctions decreases as $a\sigma \rightarrow 1$.

\section{Solution of Radial Dirac Equation}

In the radial equation independent variable $r$ is extended 
from $0$ to $\infty$. For mathematical simplicity
we change the independent variable $r$ to $r_*$ as
$$
r_{*} = r + \frac{2M r_+ + am/\sigma} {r_+ - r_-} {\rm log}
\left({r \over r_+} - 1\right) - \frac{2M r_- + am/\sigma} {r_+ - r_-} {\rm log}
\left({r \over r_{-}} - 1\right)
\eqno{(42)}
$$
(for $r > r_{+}$), here in new $r_*$ co-ordinate system horizon $r_+$ is
shifted to $-\infty$ unless $\sigma \le -\frac{am}{2Mr_+}$ [3],
so the region is extended from $-\infty$ to $\infty$.
We also choose $R_{-{1\over 2}}=P_{-{1\over 2}}$, $\Delta^{1\over 2} 
R_{+ {1 \over 2}} = P_{+ {1 \over 2}}$. Then we are defining
$$
(\lambda \pm i m_p r) = exp ({\pm i \theta}) \surd ({\lambda}^{2} + m_p^2 r^2)
$$
and 
$$
P_{+ {1 \over 2}} = \psi_{+ {1 \over 2}}\  exp\left[-{1 \over 2} i\  tan^{-1} \left({{m_p r}
\over \lambda}\right)\right],
$$
$$
P_{- {1 \over 2}} = \psi_{- {1 \over 2}}\  exp\left[+{1 \over 2} i \ tan^{-1} \left({{m_p r}
\over \lambda}\right)\right].
$$
Finally choosing,
$$
Z_{\pm} = \psi_{+ {1 \over 2}} \pm \psi_{-{1 \over 2}}
$$
and combining the differential equations (29) we get,
$$
\left(\frac{d} {d{\hat r}_*} - W\right) Z_+ = i \sigma Z_- ,
\eqno{(43a)}
$$
and
$$
\left(\frac{d} {d{\hat r}_*} + W\right) Z_- = i \sigma Z_+ ,
\eqno{(43b)}
$$
where, 
$$
{\hat{r}}_*=r_*+\frac{1}{2\sigma}tan^{-1}\left(\frac{m_pr}{\lambda}\right)
$$
and
$$
W = \frac{\Delta^{1 \over 2} (\lambda^{2} + m_p^2 r^2)^{3/2}}{\omega^2(\lambda^2+m_p^2r^2)
+\lambda m_p \Delta/2\sigma}.
\eqno{(44)}
$$
where, $\omega^2=\frac{K}{\sigma}$.

Now decoupling equations (43a-b) we get,
$$
\left(\frac{d^2} {{d {\hat r}_*}^2} + \sigma^2\right) Z_\pm = V_\pm Z_\pm .
\eqno{(45)}
$$
where, $V_{\pm} = W^{2} \pm {dW \over d\hat{r}_{*}}$ and ${\hat{r}}_*$ is
extended from $-\infty$ (horizon) to $+\infty$.

The equation (45) is nothing but one dimensional Schr\"odinger 
equation [10] with potentials $V_\pm$
and the energy of the particle $\sigma^2$ (since the system is dimensionless)
in Cartesian co-ordinate system. The equation (45) can be solved by WKB approximation
method [10-11]. The corresponding solution is [6-7],
$$
Z_{\pm}=\frac{A_{\pm}}{\sqrt{k_\pm}}exp(i{u_\pm})\pm\frac{B_{\pm}}{\sqrt{k_\pm}}
exp(-i{u_\pm})
\eqno{(46)}
$$
where,
$$
k_{\pm}=\sqrt{(\sigma^2-V_{\pm})},
\eqno{(47)}
$$
and
$$
u_\pm=\int{k_{\pm} d{{\hat{r}}_*}} .
\eqno{(48)}
$$
Now we improve the solution by introducing space dependences on coefficients
$A_\pm$ and $B_\pm$ [6-7] (this is beyond WKB approximation, because WKB deals with
solutions with constant coefficients). It is seen that far away from a black hole,
potential varies very slowly. Thus, in those regions one can safely  write,
$$
A_\pm - B_\pm = {\rm Constant} (=c).
\eqno{(49)}
$$
Since the sum of reflection and transmission coefficients must be unity,
$$
A^2_\pm+B^2_\pm=k_\pm.
\eqno{(50)}
$$
Near the horizon it is seen that potential height reduces to zero so
the reflection in that region is almost zero and transmission is almost 100\%.
This is the {\it inner boundary condition}.
Solving (49) and (50) we get analytical expression of space dependent 
reflection and transmission coefficients far away from the 
black hole which satisfy {\it outer boundary condition}. 
Combining the inner and outer boundary conditions,
we get analytical expression of space dependent coefficients $A_\pm$ and $B_\pm$ 
which is valid in whole region ($-\infty$ to $+\infty$). For details see [6-7]. 
The space dependency of $A_\pm$ and $B_\pm$ i.e. the 
transmission and reflection coefficients arises due to the variation of
potential with distance. So from the analytical expressions one can easily find
out at each point what fraction of incoming matter is going inward
and what other fraction is going outward as a result of the interaction with the black hole.
These space dependent transmission and reflection coefficients are given below [6-7],
$$
T_\pm = \frac{a^2_\pm}{k_\pm}=\frac{(c_1 + \frac{c}{2})}{h_\pm} \left(c_1 + \frac{c}{2} +
\sqrt{2 k_\pm - c^2}\right) + \frac{2 k_\pm - c^2}{4 h_\pm}
\eqno {(51a)}
$$
$$
R_\pm = \frac{b^2_\pm}{k_\pm}=\frac{(c_2 - \frac{c}{2})}{h_\pm} \left(c_2 - \frac{c}{2} +
\sqrt{2 k_\pm - c^2}\right) + \frac{2 k_\pm - c^2}{4 h_\pm}.
\eqno{(51b)}
$$ 
Here, $a_\pm$ and $b_\pm$ are defined as
$$
a_{\pm} = \frac{A_{\pm}}{\sqrt{h_\pm/k_\pm}},
\eqno{(52a)}
$$
$$
b_{\pm} = \frac{B_{\pm}}{\sqrt{h_\pm/k_\pm}}
\eqno{(52b)}
$$
which are transmitted and reflected amplitudes of the solution with modified
WKB method (going beyond WKB method) and
$$
h_\pm=\left(c_{1} + \frac{c}{2}\right)^2 + \left(c_{2} - \frac{c}{2}\right)^2
+ (c_{1} + c_{2})\sqrt{2k_\pm - c^2} + \frac{(2k_\pm - c^2)}{2},
\eqno{(53)}
$$
where, $c_1$ and $c_2$ are two constants introduced to satisfy the inner
boundary condition. The final form of the solution is
$$
Z_{\pm}=\frac{a_{\pm}}{\sqrt{k_\pm}}exp(i{u_\pm})\pm\frac{b_{\pm}}{\sqrt{k_\pm}}
exp(-i{u_\pm}) .
\eqno{(54)}
$$
Since the relation between $Z_{\pm}$ and $R_{\pm{1\over 2}}$ is known, one can
easily calculate the radial wave function $R_{\pm{1\over 2}}$.

\section{Conclusions}

In this review we write the Dirac equation in curved space-time and particularly
in Kerr geometry. From this, the behaviour of non-integral spin particles can be 
studied in curved space-time. From the form of the equation and its solution it is 
clear that in curved space the particles behave in differently than in a flat space-time.
The Newman-Penrose formalism is used to write the equation
where the basis system is null. Dirac equation is separated into angular
and radial parts. Similar separation can be possible on the background of
Dyon black hole [12]. The solution of angular component of the Dirac equation 
is first reviewed. The exact solution is possible 
for $\frac{m_p}{\sigma}=1$, otherwise the solution is approximate [5].
Unlike in the case of a Kerr black hole, the solution of the
angular equation around a Schwarzschild black hole is independent of
the azimuthal or meridional angles [5, 13,14]. This is expected because of
symmetry of the space-time.

The radial Dirac equation is solved using WKB approximation more clearly
modified WKB approximation [6-7], where the space 
dependent transmission and reflection
coefficients are calculated. Although WKB method is an approximate method,
it is improvised in such a way that spatial dependence of the coefficients of 
the  wave function is obtained. This way we ensure that the analytical solution 
is closer to the exact solution. The reflection and transmission coefficients 
were found to distinguish strongly the solutions
of different rest masses and different energies.
The solution might be of immense use in the study of the spectrum of 
particles emitted from a black hole horizon (Hawking radiation). 

\section{Acknowledgment}

It is a great pleasure to the author to thank Prof. Sandip K. Chakrabarti
for many serious discussion and giving me the opportunity to write this review.

\end{document}